\documentstyle[12pt]{article}
\normalbaselines
\begin{document}
\bibliographystyle{unsrt}

\vbox {\vspace{5mm}}

\begin{center}
{\huge\bf Classical States via Decoherence}\\[15mm]
Gh.--S. Paraoanu\footnote{On leave  from the {\it Department of Theoretical Physics, Institute of Atomic Physics, Bucharest-Magurele, PO Box MG-6, Romania}. Electronic address: paraoanu@physics.uiuc.edu  .}
 \\
{\it Department of Physics, University of Illinois at Urbana-Champaign, \\ 1110 W. Green St., Urbana, IL 61801, USA}\\[2mm]

H. Scutaru\footnote{Electronic address: scutaru@theor1.ifa.ro .}  \\
 
{\it Department of Theoretical Physics, Institute of
Atomic Physics\\ Bucharest-Magurele, POB MG-6, Romania}\\[5mm]

\end{center}

\vspace{2mm}

\begin{abstract}
 The initial states which minimize the predictability loss for a damped harmonic oscillator are identified as quasi-free states with a symmetry dictated by the environment's diffusion coefficients. For an isotropic diffusion in phase space, coherent states (or mixtures of coherent states) are selected as the most stable ones.

\end{abstract}
\vspace{1cm}
\vspace{1cm}

The emergence of classical reality from the underlying quantum description of the world is one of the most fascinating unsolved problems of present-day physics. Decoherence was proposed as a mechanism for the selection of classical (preferred) states: due to the interaction with an environment (external degrees of freedom) classical states are singled out as the most stable ones.

Zurek, Habib and Paz \cite{z} invented a criterion, called ``predictability sieve'', for distinguishing the preferred set of states from the rest of the Hilbert space: classical states are characterized by the least increase in entropy. They addressed this problem in the context of the Caldeira-Leggett model (a particle moving in a potential and linearly coupled with a bath of harmonic oscillators, \cite{z,cl,uz,zp}). They found out that coherent states are selected, via predictability sieve, as the most ``classical'' ones. 

 In contradistinction to this approach,
we will study the problem of classicality in the 
framework of Lindblad's theory,
in which the structure of the master equation is derived from very
general assumptions concerning the mathematical description of
the evolution of an open quantum system \cite{l}. This strategy leads to general results (since Lindblad equations work for a large class of physically interesting systems \cite{ss}), and also avoids any problems related to the non-preservation of the positivity of the reduced density matrix, which was perceived by some authors \cite{am} as an inconvenient of the Caldeira-Leggett type of evolution.
We will identify the states with classical behavior as the states which minimize the rate of entropy increase. We will apply this criterion in an identical fashion for both pure and mixed initial states, using a formalism that doesn't require one to discriminate between them.

Lindblad's result shows that a general form for the generator of
a completely positive dynamical semigroup in the Schr\"{o}dinger
picture is \cite{l} 

\begin{equation}
L(\rho)=-\frac{i}{\hbar}[H,\rho]+\frac{1}{2\hbar}
\sum_{j}([V_{j}
\rho,V_{j}^{+}]+[V_{j},\rho V_{j}^{+}]),\label{l}
\end{equation}
where for the operators $V_{j}$, we will take \cite{ss}
$
V_{j}=a_{j}p+b_{j}q,\label{ec}
$
where $j=1,2$ while $a_{j}$ and $b_{j}$ are complex numbers. For
the Hamiltonian $H$ of the system 
we will consider \cite{ss} a second-order polynomial operator in
coordinate and position written as a sum of a harmonic-oscillator
part and a mixed term
\begin{equation}
H=\frac{m\omega^{2}}{2}q^{2} + \frac{p^{2}}{2m} + {\mu\over 2}(pq+qp).
\end{equation}
We denote the diffusion coefficients by
\begin{equation}
D_{qq}={\hbar\over2}\sum_{j=1}^{2}|a_{j}|^{2},~~D_{pp}={\hbar\over2}\sum
_{j=1}^{2}|b_{j}|^{2},~~D_{pq}=D_{qp}=-{\hbar\over2}Re\sum_{j=1}^{2}
a_{j}^{*}b_{j},
\end{equation}
and the friction constant is
\begin{equation}
\lambda=-Im\sum_{j=1}^{2}a_{j}^{*}b_{j}.
\end{equation}
 It is easy to show  \cite{ss} that the following inequality
must be satisfied
\begin{equation}
D_{pp}D_{qq}-D_{pq}^{2}\geq{\lambda^{2}\hbar^{2}\over 4}.\label{s}
\end{equation}
 With the
above notation, the master equation which governs the evolution
of the system  takes the form
 \begin{eqnarray}
\frac{d\rho(t)}{dt}&=&-{i\over\hbar}[H,\rho(t)]
-{i\lambda\over\hbar}([q,\rho(t) p]-[p,\rho(t)
q]) \nonumber\\ & &
-{D_{qq}
\over\hbar^2}[p,[p,\rho(t)]]
-{D_{pp}\over\hbar^2}[q,[q,\rho(t)]] 
+{2D_{pq}\over
\hbar^2}[p,[q,\rho(t)]]
.\label{ll} \end{eqnarray}

For the correlations between two operators $C$ and $C'$ we will
use the definition
\begin{equation}
\sigma_{C,C'}(t)=Tr\left(\rho(t){{CC'+C'C}\over 2}
\right)-\sigma_{C}(t)\sigma_{C'}(t)
,\end{equation} where $\sigma_{C}(t)=Tr(\rho(t) C)$ and $\sigma_{C'}(t)=Tr(\rho(t) C')$. $\Sigma(t)$ will denote the dispersion matrix
\begin{equation}\Sigma(t)
=\left(\matrix{m \omega \sigma_{qq}(t)&\sigma_{pq}(t)\cr\sigma_{pq}(t)&
{\sigma_{pp}(t)\over m \omega}\cr}\right).
\end{equation}
With the notations above, Heisenberg's inequality reads 
~$det\Sigma(t)\geq
{\hbar^2\over4}$.~
For $\Sigma(t)$ the time-evolution is known \cite {ss}
\begin{equation}
{d\Sigma(t) \over dt}=Y\Sigma(t)+\Sigma(t) Y^{T}+2\cal{D}\label{ev},
\end{equation}
where
\[
Y=\left(\matrix{-(\lambda-\mu) & \omega\cr-\omega &
-(\lambda+\mu)\cr}\right),\]
$Y^{T}$ is the transposed matrix of $Y$ 
and $\cal D $ is a $2\times 2$ matrix with elements ${\cal D}_{qq}=m\omega 
D_{qq}$, ${\cal D}_{pq}={\cal D}_{qp}=D_{qp}$ and
${\cal D}_{pp}={D_{pp}\over m\omega}$.

For the case of the damped harmonic oscillator, it is known that
a certain class of states (quasi-free states \cite{s}) has the
property of invariance under the action of the quantum semigroup.
This can be seen by writing the Fokker-Planck equation which corresponds
to (\ref{ll}) in the form \cite{i}
\begin{equation}
{\partial f_{W}(x_{1}, x_{2}, t)\over\partial t}=
\sum_{i,j=1}^{2}Y_{ij}{\partial\over 
\partial x_{i}}
(x_{j}f_{W}(x_{1}, x_{2}, t))
+{1\over 2\hbar}\sum_{i,j=1}^{2}{\cal D}_{ij}{\partial^{2}\over\partial x_{i}
\partial x_{j}} f_{W}(x_{1}, x_{2}, t),\label{f}
\end{equation}
where $f_{W}$ is the Wigner function of a quasi-free state 
\begin{equation}
f_{W}(x_{1}, x_{2}, t)=
[(2\pi)^2det\Sigma(t)]^{-{1\over 2}}\exp\left\{-{1\over2}X^{T}(t)\Sigma^{-1}X(t)
\right\},\label{w}
\end{equation}
and $X(t)=\left(\matrix{x_{1}-\sigma_{q}(t)\cr x_{2}-\sigma_{p}(t)}\right)$. It is now
easy to verify that the Gaussian Wigner function
(\ref{w}), with
the time-dependence of the mean values $\sigma_{p}(t)$, $\sigma_{q}(t)$  and of dispersions 
$\sigma_{qq}(t)$, $\sigma_{pp}(t)$ , $\sigma_{pq}(t)$  given by
equations (3.26 - 3.27) from \cite{ss},
is a solution of the equation (\ref{f}). Thus, quasi-free states
are preserved during the evolution of the system.

Starting with such a state, we are interested to calculate the rate of linear
entropy variation. The initial states with the lowest rate of linear entropy increase will be identified as the most stable ({\it i.e.} the most classical-like) states.

The linear entropy is a convenient measure of the purity of a quantum state and is defined by
\begin{equation} {\it s}(\rho) = 1 - Tr(\rho ^2).\end{equation}
For a quasi-free state we have 
\begin{equation} {\it s}(\rho) = 1 - {1 \over A }. \end{equation}
where $A(t)$ is the ``area'' in phase space,
$A(t)={2\over\hbar}\sqrt{det\Sigma(t)}$.
The condition for a quasi-free state to be a pure one is $A(t)=1$, that is
$det\Sigma(t)={\hbar^{2}\over4}$ (the equality case in the Heisenberg 
relation). The
time-derivative of $s(t)$ can be calculated by making use of the following 
relation 
\begin{equation} {d(\ln det\Sigma(t)) \over dt}~=~Tr \left({d \Sigma(t) \over
 dt} \Sigma ^{-1}(t) \right),\label{ff} \end{equation}
where
\begin{equation} \Sigma ^{-1}(t)~=~{1 \over det\Sigma(t)} \left(
\matrix{{\sigma_{pp}(t)\over m\omega} & -\sigma_{pq}(t)
\cr -\sigma_{pq}(t) & m\omega\sigma_{qq}(t) \cr} \right). \end{equation}
Then we obtain from (\ref{ev}) and (\ref{ff}) that
\begin{equation} {dA(t) \over dt}~=~{{2 \over \hbar} \sqrt{det\Sigma(t)}} 
( TrY~+~Tr({\cal D}\Sigma ^{-1}(t))). \label{eon}\end{equation}

But the rate of the linear entropy increase is given by
\begin{equation} {d{\it s(t)} \over dt} = {1\over A(t)^2} 
{dA(t)\over dt},\label{linentr}\end{equation}
so  the behavior of the rate of linear entropy increase
 is given entirely by the time derivative (\ref{eon}) of the area $A(t)$  for
all states (including pure states). In the following we shall
find the squeezing parameter of the initial state for which the rate of 
increase of $A(t)$ is minimized. 

Let us notice, for the beginning, that
the positive $2\times 2$ 
matrix $\Sigma(t)$ can be diagonalized, at any particular instant $t$,
 in the form (see also
\cite{s1,b})
\begin{equation}
\Sigma(t)={\hbar A(t)\over 2}O^{T}(t)\left(\matrix{\aleph^{2}(t)&0\cr 0&
\aleph^{-2}(t)}\right)O(t).\label{dec}
\end{equation}
where $\aleph (t)$ 
is a real positive number (the squeezing parameter) $A(t)$ is the area 
occupied by the system in phase space and
$O(t)$ is an orthogonal symplectic matrix
for which we will employ the usual form
\[ O(t)=\left(\matrix{\cos\theta(t)&-\sin\theta(t)\cr\sin\theta(t)&
\cos\theta(t)\cr}\right).
\]

A similar formula holds for the ${\cal D}$-matrix
\begin{equation}
{\cal D}=\frac{\hbar\Delta}{2}O_{D}^{T}\left(\matrix{d^{2}&0\cr 0&
d^{-2}}\right)O_{D},
\end{equation}
with 
\begin{equation}
O_{D}=\left(\matrix{\cos\varphi &-\sin\varphi \cr\sin\varphi &
\cos\varphi \cr}\right).
\end{equation}
Here, $\Delta$ is a parameter controlling the intensity of diffusion, $d$ characterizes the degree of anisotropy and $\varphi$ is the rotation angle.

Now, (\ref{eon}) and (\ref{linentr}) imply
\begin{equation}
\frac{ds}{dt}\begin{array}{|c}  \\ t=0 \end{array} = \frac{1}{A(0)}\left[ -2\lambda +Tr\left(\Sigma^{-1}(0){\cal D}\right)\right].\label{yeye}
\end{equation}

This result is
even more general when the system is in a pure state at t=0 (so that $A(0)=1$), in the sense that it does not depend on the kind
of initial states we are starting with (quasi-free states or not).
Indeed, in  \cite{ss}
it was shown that 
\begin{equation} {d Tr(\rho ^2) \over dt} = 2 Tr(\rho L(\rho)) =
{2 \over \hbar} \sum_{j}(Tr(\rho V_{j} \rho V_{j}^{*})-
Tr(\rho ^2 V_{j}^{*}V_{j})) \geq 0 .\end{equation}
For a pure state $\rho ^2 = \rho$ and
$\rho O \rho = Tr(\rho O) \rho$ for any selfadjoint
operator $O$. Then
\begin{equation} {dTr(\rho ^2) \over dt} = {2 \over \hbar}
\sum_{j} (\vert Tr(\rho V_{j})\vert ^2 -
Tr(\rho V{j}^{*}V_{j})) \geq 0 .
\end{equation}
We have $Tr(\rho V_{j}) = a_{j} \sigma_{p} + b_{j}\sigma_{q}$.
Hence
we get
\begin{eqnarray}\nonumber
{d\over dt}(1-Tr(\rho^{2}))\begin{array}{|l}\\t=0\end{array}
&=&-2\lambda+{4\over\hbar^{2}}(D_{qq}\sigma_{pp}(0)
+D_{qq}\sigma_{pp}(0)-2D_{pq}\sigma_{pq}(0))\\ \nonumber
&=&-2\lambda +Tr\left(\Sigma^{-1}(0)\cal{D}\right),
\end{eqnarray}
which has the same form as the rate of linear entropy increase calculated
before (see (\ref{yeye})) 
for pure ($A(0)=1$) initial quasi-free states.

We are interested in finding the states which produce the least 
increase of the phase-space area at
the initial moment $t=0$. By minimizing the expression 
\begin{eqnarray} \nonumber
-2\lambda +Tr\left(\Sigma^{-1}(0)\cal{D}\right)&=& -2\lambda + \frac{\Delta}{A(0)}\{\cos^{2}(\theta (0) -\varphi )[\aleph^{2}(0)d^{-2}+\aleph^{-2}(0)d^{2}] \\ \nonumber
 & &+\sin^{2}(\theta (0) -\varphi )[\aleph ^{2}(0)d^{2}+\aleph^{-2}(0)d^{-2}]\},\nonumber
\end{eqnarray}
 
with respect to $\aleph (0)$ and $\theta(0)$,
one gets
\begin{equation}
min\left[\frac{ds}{dt}\right]\begin{array}{|c}  \\ t=0 \end{array}
=2\frac{\Delta - A(0)\lambda }{A(0)^{2}} .
\end{equation}
corresponding to
\begin{equation}
\begin{array}{c}\aleph^{*}(0)=d ,\\ \theta^{*}(0)=\varphi .\end{array}
\end{equation} 

So, in the general case of an anisotropic diffusion, the minimum variation of the area in phase space is obtained when the squeezing parameter of the state equals d, the degree of anisotropy of the diffusion, and the characteristic rotation angle of the correlation and diffusion matrices are equal. For an isotropic diffusion, ${\cal D}_{pp}={\cal D}_{qq}$, ${\cal D}_{pq}=0$, {\it i.e.} d=1
(and the rotation angle vanishes trivially from all the relations, since now the system has rotation symmetry in phase space); we obtain $\aleph^{*}(0)=1$. This case corresponds to many models of dissipation, especially from quantum optics (see \cite{ik}); the same result was obtained by Zurek, Habib and Paz \cite{z} in the context of the Caldeira-Leggett environment model.
In other words, a phase-space isotropic environment favors a symmetric state, while an anisotropic environment selects states with the same degree of anisotropy.

 When the initial state is pure, 
\begin{equation}
min\left[\frac{ds}{dt}\right]\begin{array}{|c}  \\ t=0 \end{array}
=2\left(\Delta - \lambda\right)\geq 0, 
\end{equation}
where the last inequality comes from (\ref{s}) and expresses the fact that the
linear entropy of pure states in an environment always increases, because the pure initial states become more and more mixed.

Our result shows that the values of $\aleph^{*}(0)$ and 
$\theta^{*}(0)$ are independent of the overall magnitude $\Delta$ of the diffusion. They are also independent of $A(0)$. Thus, the same degree of squeezing is singled out, irrespective of the  purity of the initial state, thus confirming previous insights \cite{z,zp} regarding the structure of the mixed preferred states: they can be seen as the thermalization of the selected pure states.   
   
We conclude by emphasizing the main results of this paper. In
general, the pure or mixed state which produces the minimum rate
of increase in the area occupied by the system in phase--space
is a quasi-free state which has the same symmetry
as that induced on the evolution in phase-space by the
diffusion coefficients. For isotropic phase-space diffusion, the selected pure states are the coherent states.

\end{document}